\documentclass[aps,prl,twocolumn,showpacs,superscriptaddress,groupedaddress,reprint]{revtex4}  % for review and submission

\usepackage{graphicx}  % needed for figures
\usepackage{dcolumn}   % needed for some tables
\usepackage{bm}        % for math
\usepackage{amssymb}   % for math

\usepackage{xspace}
\usepackage{amsmath}
\usepackage{abbrevs}
\usepackage[]{units}	%Auto-formats units
\usepackage{color}

\usepackage{float} 
\usepackage{wrapfig} 

\newcommand{\bra}[1]{\langle #1 |}
\newcommand{\ket}[1]{| #1 \rangle}

\newabbrev\SPDC{spontaneous parametric downconversion (SPDC)}[SPDC]
\newabbrev\APD{avalanche photodiode (APD)}[APD]
\newabbrev\TBP{time-bandwidth product (TBP)\xspace}[TBP]
\newabbrev\FWM{four-wave mixing (FWM)\xspace}[FWM]
\newabbrev\SNR{signal-to-noise ratio (SNR)\xspace}[SNR]
\newabbrev\LOQC{linear optical quantum computing (LOQC)\xspace}[LOQC]
\newabbrev\PBS{polarizing beam-splitter (PBS)}[PBS]
\newabbrev\LMBS{{\em light-matter beam-splitter} (LMBS)\xspace}[LMBS]
\newabbrev\BBO{$\beta$-Barium Borate (BBO)}[BBO]
\newabbrev\HOMI{Hong--Ou--Mandel (HOM)\xspace}[HOM]
\newabbrev\OPO{optical parametric oscillator (OPO)\xspace}[OPO]
\makeatletter
\renewcommand\maybe@space@{%
  \maybe@ictrue % <= this is new
  \expandafter   \@tfor
    \expandafter \reserved@a
    \expandafter :%
    \expandafter =%
                 \nospacelist
                 \do \t@st@ic
  \ifmaybe@ic % <= this is new
    \space
  \fi
}

\begin{document}

% The following information is for internal review, please remove them for submission
%\widetext
%\leftline{Version 1 as of \today}
%\leftline{Primary authors: Duncan G. England}
%\leftline{To be submitted to PRL}
%\leftline{Comment to {\tt d0-run2eb-nnn@fnal.gov} by xxx, yyy}

% the following line is for submission, including submission to the arXiv!!
%\hspace{5.2in} \mbox{Fermilab-Pub-04/xxx-E}

%\title{Demonstration of high-bandwidth qubit storage in a diamond quantum memory}
%\title{A room-temperature quantum memory for ultrafast photonic qubits}
%\title{Non-classical storage in room-temperature diamond}
%\title{Storage of THz-bandwidth photonic qubits in room-temperature diamond}
\title{Storage of polarization-entangled THz-bandwidth photons in a diamond quantum memory}

\author{Kent~A.G.~Fisher}\affiliation{Institute for Quantum Computing and Department of Physics \& Astronomy, University of Waterloo, 200 University Avenue West, Waterloo, Ontario N2L 3G1, Canada}

\author{Duncan~G.~England} \affiliation{National Research Council of Canada, 100 Sussex Drive, Ottawa, Ontario, K1A 0R6, Canada}

\author{Jean-Philippe~W.~MacLean}\affiliation{Institute for Quantum Computing and Department of Physics \& Astronomy, University of Waterloo, 200 University Avenue West, Waterloo, Ontario N2L 3G1, Canada}

\author{Philip~J.~Bustard} \affiliation{National Research Council of Canada, 100 Sussex Drive, Ottawa, Ontario, K1A 0R6, Canada}

\author{Khabat~Heshami} \affiliation{National Research Council of Canada, 100 Sussex Drive, Ottawa, Ontario, K1A 0R6, Canada}

\author{Kevin~J.~Resch}\affiliation{Institute for Quantum Computing and Department of Physics \& Astronomy, University of Waterloo, 200 University Avenue West, Waterloo, Ontario N2L 3G1, Canada}

\author{Benjamin~J.~Sussman} \email{ben.sussman@nrc.ca}\affiliation{National Research Council of Canada, 100 Sussex Drive, Ottawa, Ontario, K1A 0R6, Canada}\affiliation{Department of Physics, University of Ottawa, Ottawa, Ontario, K1N 6N5, Canada}

\date{\today}

\begin{abstract}

\noindent Bulk diamond phonons have been shown to be a versatile platform for the generation, storage, and manipulation of high-bandwidth quantum states of light.  Here we demonstrate a diamond quantum memory that stores, and releases on demand, an arbitrarily polarized $\sim$250\,fs duration photonic qubit.  The single-mode nature of the memory is overcome by mapping the two degrees of polarization of the qubit, via Raman transitions, onto two spatially distinct optical phonon modes located in the same diamond crystal. The two modes are coherently recombined upon retrieval and quantum process tomography confirms that the memory faithfully reproduces the input state with average fidelity $0.784\pm0.004$ with a total memory efficiency of $(0.76\pm0.03)\%$. In an additional demonstration, one photon of a polarization-entangled pair is stored in the memory. We report that entanglement persists in the retrieved state for up to 1.3\,ps of storage time.  These results demonstrate that the diamond phonon platform can be used in concert with polarization qubits, a key requirement for polarization-encoded photonic processing.

\end{abstract}
 
%\pacs{42.50.Ex, 03.67.Hk, 81.05.ug}
\maketitle

\noindent  Extreme light-matter interactions continue to drive new photonic developments.  At one end of the intensity scale, modern ultrafast lasers produce extremely intense pulses of light, which can be used to observe fundamental non-linear optical effects such as harmonic generation, wave mixing, and stimulated Raman scattering~\cite{Boyd2003}. At the other end of the intensity scale, exquisite control of the light-matter interface enables interaction between a single photon and a single atom~\cite{Wilk2007}.  At the confluence of these two fields --- nonlinear optics and quantum optics  ---  intense optical pulses in nonlinear media can be used to control quantum systems for the generation~\cite{Eisaman2011}, storage~\cite{Kozhekin2000}, and processing~\cite{McGuinness2010} of single photons.  In this context, bulk diamond is emerging as a versatile room-temperature platform for both nonlinear optics ~\cite{Lux2016,Reilly2015} and quantum optics~\cite{Lee2012, Hou2016} due to its high optical nonlinearity, large phonon energy, simple energy level structure and unrivalled mechanical strength. 

Recent work has shown that diamond can be used as an absorptive quantum memory, storing THz-bandwidth single photons and releasing them, on demand, several picoseconds later~\cite{England2013,England2015}. Absorption, and retrieval, of single photons is achieved via a Raman transition mediated by intense ultrafast laser pulses, thereby leveraging both the quantum and nonlinear optical benefits of the diamond platform. Beyond storage, the diamond memory has demonstrated fundamental signal processing steps including frequency and bandwidth manipulation of single photons~\cite{Fisher2016} and beam-splitter operations~\cite{England2016}. In previous work, the memory has only been able to store photons in a single optical mode and thus has been unable to store quantum information where $\ket{0}$ and $\ket{1}$ are encoded in different optical modes. Multi-mode storage is therefore a key outstanding challenge in the development of the diamond quantum memory. In this work we describe a two-mode diamond memory that is capable of storing both horizontal and vertical polarization and is therefore well-suited to interfacing with polarization qubits.

%%% FIG %%%
\begin{figure*}
\center{\includegraphics[width=1.0\linewidth]{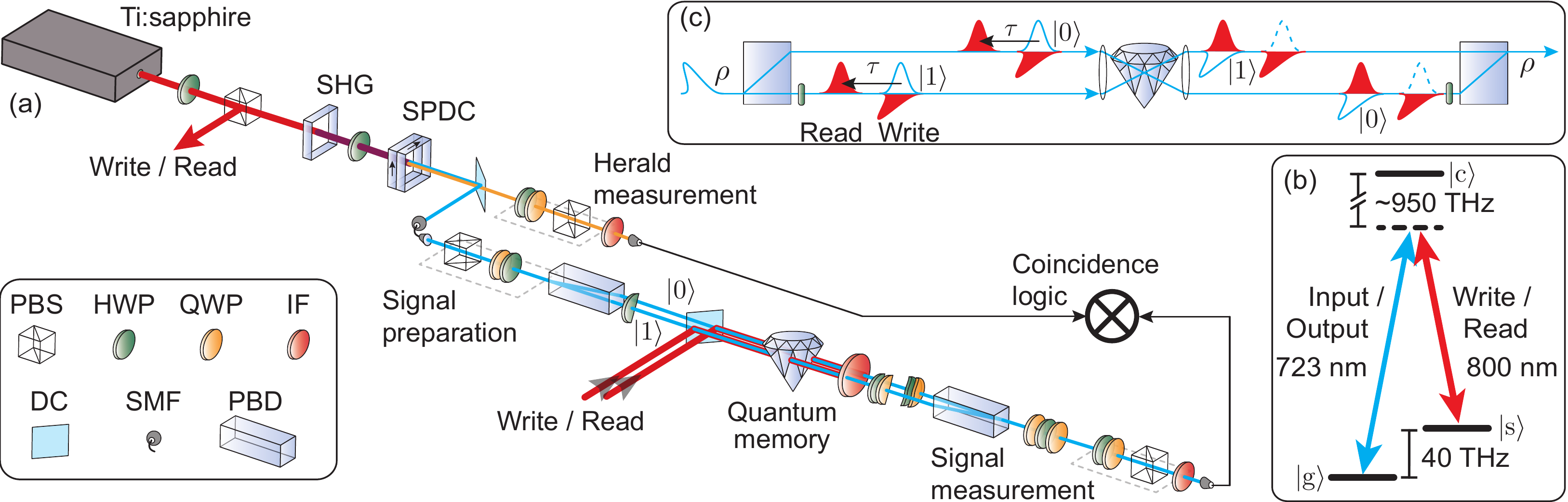}}
\caption{(a) Experimental setup. Pulses for the read, write and photon source pump originate from a Ti:sapphire oscillator. Photon pairs (signal, herald) are generated via SPDC pumped by the second-harmonic (SHG) of the laser and spatially filtered in single-mode fibre (SMF). A polarization qubit is prepared in the signal arm (blue), and separated into spatial modes, $\ket{0}$ and $\ket{1}$, by a polarizing beam displacer (PBD) and sent to the memory. (For entangled photon storage the signal preparation step is removed.) Read and write pulses are prepared in a displaced Mach-Zehnder and Michelsen interferometers (not shown), and are overlapped with $\ket{0}$ and $\ket{1}$ modes on a dichroic mirror (DC). After retrieval, half- (HWP) and quarter-wave plates (QWP) in each mode correct polarization, the two spatial modes are recombined using a second beam displacer, and phase rotations are corrected with a QWP-HWP-QWP combination. The polarization is analyzed by a HWP, QWP and polarizing beamsplitter (PBS). After an interference filter (IF) coincident detections with the herald mode are counted.  (b) $\Lambda$-level diagram for memory process in diamond. Input and write fields transition the crystal from its ground state ($\ket{\text{g}}$) to the optical phonon level ($\ket{\text{s}}$); read and output fields drive the reverse the transition. Both frequencies are far detuned from the conduction band ($\ket{\text{c}}$). (c) A polarization qubit $\rho$ is converted to spatial modes $\ket{0}$ and $\ket{1}$ by a beam displacer. Each mode is focussed onto the diamond and stored for time $\tau$ in the diamond memory. After retrieval the two modes are recombined to form a polarization qubit. 
}
\label{fig:Concept}
\end{figure*}

It is not possible to store a polarization qubit in a single spatial mode in diamond because both horizontal and vertical polarizations couple to the same optical phonon mode, so independent readout is impossible. Similarly time-bin or frequency-encoded quantum information could not be faithfully stored; therefore, like many other on-demand quantum memories, the diamond memory is intrinsically single-mode~\cite{Choi2008, Reim2010, Zhang2011}. Instead, in order to store a polarization qubit, we employ two independent memories located inside the same diamond crystal. The polarization qubit is split into its constituent horizontally- and vertically-polarized components and each is stored in a different spatial mode~\cite{Cho2010,England2012,Kupchak2015,Parigi2015,Gundogan2012}. After retrieval from the memory, the two modes are recombined to recover the original qubit. We use our two-mode memory to store a THz-bandwidth polarization qubit, with average fidelity of 78\%. Furthermore, we store one photon of a polarization-entangled pair and achieve a fidelity of 76\% with the input entangled state. These results are an important benchmark for the diamond platform because they show that, by spatially multiplexing, we can faithfully store a two-mode quantum state. (This method can also be used to store higher-dimensional quantum states.) We therefore expect that other diamond-based quantum processing protocols such as frequency conversion~\cite{Fisher2016} and beam-splitter operations~\cite{England2016} could also be applied to polarization qubits.

%%% EXPERIMENT %%%
The memory functions as follows.  A single photon (\emph{signal}) is stored in the optical phonon modes of the diamond lattice ($\ket{\text{s}}$ in Figure~\ref{fig:Concept}(b)) via a Raman transition stimulated by a strong \emph{write} pulse. The 40\,THz splitting between ground and storage states, and the $\sim$950\,THz detuning from the conduction band ($\ket{\text{c}}$) allow for the storage of THz-bandwidth light with a quantum-level noise floor~\cite{England2013}; our input signal photons have 2.4\,THz bandwidth (4.1\,nm at centre wavelength 723\,nm). After storage for time $\tau$, the photon is retrieved from the memory by a \emph{read} pulse.

The diamond crystal was manufactured by Element Six Ltd. using chemical vapour deposition. The diamond is cut along the $\left \langle 100 \right \rangle$ face of the crystal lattice.  In this configuration Raman selection rules require that the write pulse and signal photon have orthogonal polarizations.  Similarly, the photon retrieved from the memory has orthogonal polarization to the read pulse. The polarization-sensitive nature of the protocol necessitates a low bifrefringence in the crystal. The diamond must also have a high purity to eliminate fluoresence noise from color centres.

To store a polarization qubit we use two independent memory modes, each requiring its own set of write and read pulses (see Figure~\ref{fig:Concept}(c)).  Figure~\ref{fig:Concept}(a) shows the experimental setup.  A mode-locked Ti:sapphire oscillator operating at 80\,MHz outputs 190\,fs pulses with 800\,nm centre wavelength. The laser power is divided between read/write pulses and the pump for the photon source. The source generates photon pairs by pumping type-I \SPDC in a 1\,mm \BBO crystal with the second harmonic of the pump pulse. The detection of an 895\,nm photon \emph{heralds} the presence of a 723\,nm photon in the signal mode. When the quantum memory is not active --- no control pulses are present and input photons are transmitted through the diamond --- we measure a coincidence rate of $\sim$1100\,Hz between signal and herald detections with a coincidence window of 1\,ns.  The polarization qubit is prepared on the signal photon using a \PBS, quarter- (QWP) and half-wave plate (HWP) which is then mapped into spatial modes, using a calcite polarizing beam displacer (PBD) which separates horizontal ($\ket{H}$) and vertical ($\ket{V}$) polarizations by 4\,mm in the transverse direction.  The two signal modes, $\ket{0}$ and $\ket{1}$, are overlapped spatially and temporally with their corresponding write pulses and focussed in the diamond. With 4.4\,nJ of energy in each write pulse we observe a write efficiency of $\eta_w=6.2\%$ for each memory mode.  The PBDs on either side of the diamond form an interferometer which is passively stable on the order of tens of hours~\cite{Cho2010,England2012}. However, this stability requires that both signal modes pass through the same lens, bringing them to a focus at the same spot in the diamond (Figure~\ref{fig:Concept}(c)). Geometric effects in the interferometer, i.e., the PBDs and a HWP in one arm, result in the two modes being temporally offset 16.7\,ps such that the $\ket{0}$ mode is stored, retrieved, and the crystal restored to its ground state before the $\ket{1}$ mode arrives at the diamond. This temporal delay avoids any cross-talk between the spatially overlapped memories.

After time $\tau$, two 4.4\,nJ read pulses retrieve the $\ket{0}$ and $\ket{1}$ modes from memory. The retrieved photon modes are coherently recombined on a polarizing beam displacer to restore the polarization qubit. Any light not stored in the memory has polarization orthogonal to the retrieved signal and is extinguished by the beam displacer. The retrieved polarization qubit is analyzed using an HWP-QWP-PBS combination, collected into single-mode fibre and detected on an avalanche photodiode; joint detections of signal and herald photons are recorded by coincidence counting logic. 

\begin{figure} 
\center{\includegraphics[width=0.85\linewidth]{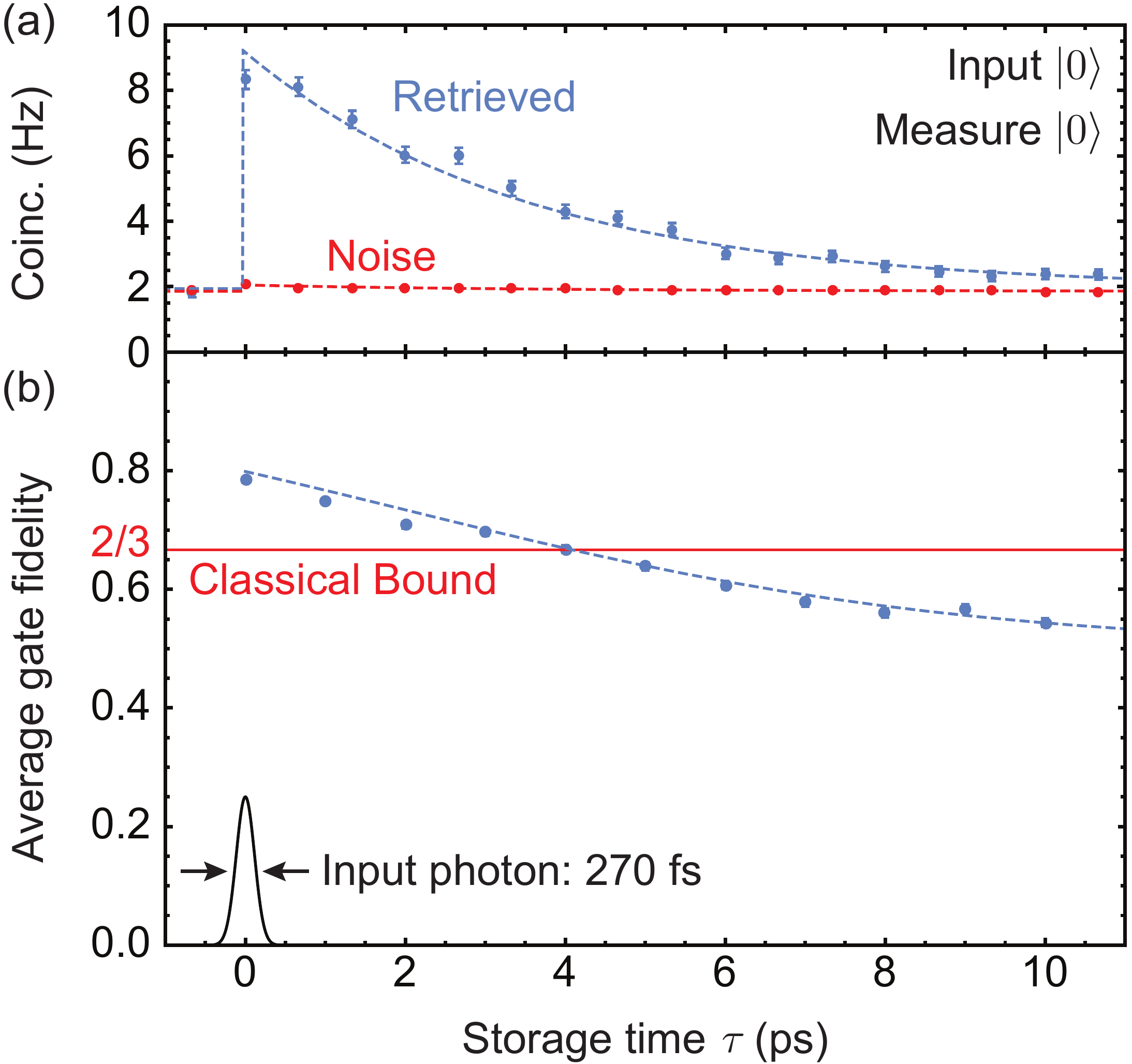}}
\caption{ (a) Measured herald-signal (blue) coincidences after retrieval, where the input state is $\ket{\text{0}}$ and the output projected onto $\ket{\text{0}}$.  Error bars show one standard deviation assuming Poissonian noise. Coincidences due to noise photons (red), i.e., when there is no input signal, are also shown. An exponential decay function $a + b e^{-\tau / \tau_\text{m}}$ (dashed) is fit to the data yielding a decay time of $\tau_\text{m} = 3.5$\,ps. (b) Average fidelity of the memory as a function of storage time.  Error bars, estimated using Monte Carlo simulations, are too small to be seen on this scale.   The memory operates above the $2/3$ bound for 3\,ps, over 11 durations of the 270\,fs input photon (shown for reference). The average fidelity for the memory model, $[1+p(\tau)]/2$, is also plotted (dashed).
}
\label{fig:AvgFid}
\end{figure}

Figure~\ref{fig:AvgFid}(a) shows the coincidence rate of retrieved signal photons as a function of storage time, $\tau$, when the qubit is prepared as $\ket{H}$, corresponding to the $\ket{0}$ spatial mode. At its peak we measure a retrieved coincidence rate of $8.3 \pm 0.3$\,Hz.  The memory efficiency is then $\eta_\text{m} = (0.76 \pm0.03)\%$, implying a retrieval efficiency of $\eta_r = \eta_\text{m} / \eta_w = 12.2\%$. Memory efficiency is limited by the amount of available laser power for write and read pulses.  
An exponential decay function, $a+b e^{-\tau / \tau_\text{m}}$,  fit to retrieval rate data returns a characteristic memory lifetime of $\tau_\text{m}=3.5$\,ps, which is limited by decay of optical phonons to acoustic phonons~\cite{Klemens1966}. 

Spurious noise processes in the memory can produce output light from the memory even when no photon is input. Comparing cases where there is, and is not, an input photon present we find a constant noise rate of $N_c = 3.737 \pm 0.004$\,Hz and a time-dependent noise rate $N_0 e^{-\tau / \tau_\text{m}}$, where $N_0 = 0.464 \pm 0.008$\,Hz; polarization analysis removes half of this produced noise.  This implies that of the retrieved coincidence rate at $\tau = 0$, a factor of $S_0 = 6.2 \pm 0.3$\,Hz is due to stored signal photons re-emitted from the memory.

%We fit the function $R = N_c / 2  + (S_0 + N_0 / 2) e^{-\tau / \tau_\text{m}}$ to the data where the fitted signal rate $S_0 = 7.1$\,Hz at $\tau = 0$, and memory lifetime is $\tau_\text{m}=3.5$\,ps, which is limited by decay of optical phonons to acoustic phonons~\cite{Klemens1966}.  Spurious noise processes in the memory can trigger counts when no photon is input to the memory. We find a constant noise rate of $N_c = 3.88$\,Hz, a time-dependent noise rate $N_0 e^{-\tau / \tau_\text{m}}$, where at $\tau = 0$ we find $N_0 = 0.38$\,Hz; polarization analysis removes half of the produced noise.  

%%% FIG %%%

%%%% Results %%%% 

%%% FIG %%%
\begin{figure} 
\center{\includegraphics[width=1.0\linewidth]{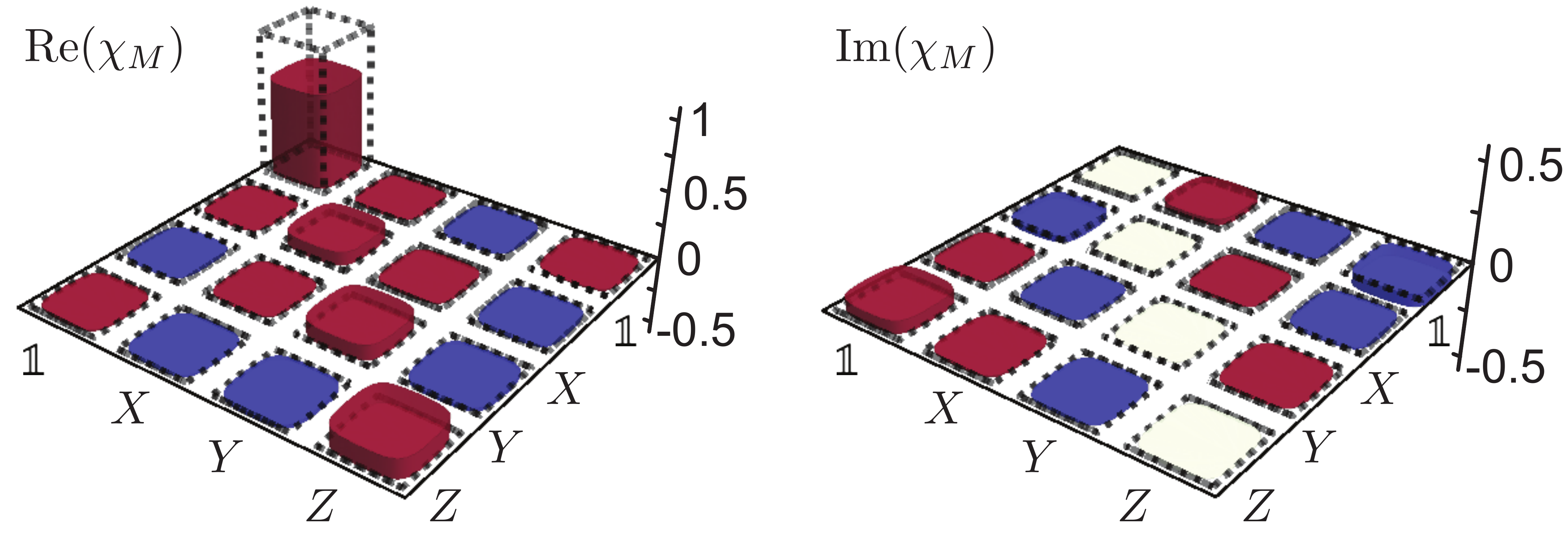}}
\caption{ Real and imaginary parts of $\chi_M$, the reconstructed process matrix for the memory in the Pauli operator basis $\{ \openone, X, Y, Z \}$. Red (blue) corresponds to positive (negative) values. The process fidelity~\cite{Jozsa1994, procfid} with the ideal memory (dashed black) is $0.677\pm 0.006$ at its peak, corresponding to an average fidelity of $0.784 \pm 0.004$. 
}
\label{fig:ProcTomo}
\end{figure}

%%% Quantum state tomography %%%
An ideal quantum memory leaves an arbitrary input  state $\rho_\text{in}$ unchanged after storage.  However, owing to noise and phonon decay the output state can be depolarized.  The output state is therefore modeled as a weighted sum of the input state $\rho_\text{in}$ and a depolarized state $\openone_2 /2$, with respective weights $p(\tau)$ and $1-p(\tau)$:
\begin{equation}
\rho_\text{out} = p(\tau) \rho_\text{in} + \left [ 1-p(\tau) \right ] \openone_2 /2,
\label{eq:channel}
\end{equation}
\noindent where $p (\tau) = S_0 e^{-\tau / \tau_\text{m}} / [ N_c + (S_0 + N_0)e^{-\tau / \tau_\text{m}} ]$ is the probability that a given photon output from the memory corresponds to successful retrieval rather than noise.  We characterize the diamond memory using quantum process tomography, which has the goal of reconstructing a process matrix $\chi_m$, such that $\rho_\text{out} = \sum_{m,n} \chi_{m,n} \sigma_m \rho_\text{in} \sigma_n$ where $\sigma_i$ are the Pauli matrices.  The process matrix $\chi_m$ for the noisy memory in Eq.~\ref{eq:channel} is diagonal in the Pauli operator basis with entries $\{ [1+3p(\tau)]/4, [1-p(\tau)]/4, [1-p(\tau)]/4, [1-p(\tau)]/4 \}$. In the experiment we input an over-complete basis set $\{ \ket{0}=\ket{H}$, $\ket{1}=\ket{H}$, $\ket{\pm}=(\ket{H} \pm \ket{V})/ \sqrt{2}$, $\ket{\pm_y}=(\ket{H} \pm i \ket{V})/ \sqrt{2} \}$ to the memory. After retrieval we measure the output in the same basis set, and perform a maximum likelihood reconstruction~\cite{Chow2009} of $\chi_\text{m}$ (real and imaginary parts of $\chi_\text{m}$ are shown in Figure~\ref{fig:ProcTomo}).  
The average fidelity~\cite{Horodecki1999}, i.e., the average of input-output state fidelities~\cite{Jozsa1994}, is $\mathcal{F}_\text{avg} = 0.784\pm0.004$ at its peak, where uncertainty is estimated by Monte Carlo simulation with Poissonian noise added to measured counts.   A classical, i.e., measure and re-send, memory can function with an average fidelity of $2/3$~\cite{Massar1995}.  As shown in Figure~\ref{fig:AvgFid}(b), our quantum memory exceeds this bound up to 3\,ps of storage which is over 11 times the input photon coherence time. The average fidelity decays with increased storage time as expected by the process in Eq.~\ref{eq:channel}, for which $\mathcal{F}_\text{avg}(\tau) = [1+p(\tau)]/2$; for non-classical storage we require $p(\tau) > 1/3$.

\begin{figure} 
\center{\includegraphics[width=0.85\linewidth]{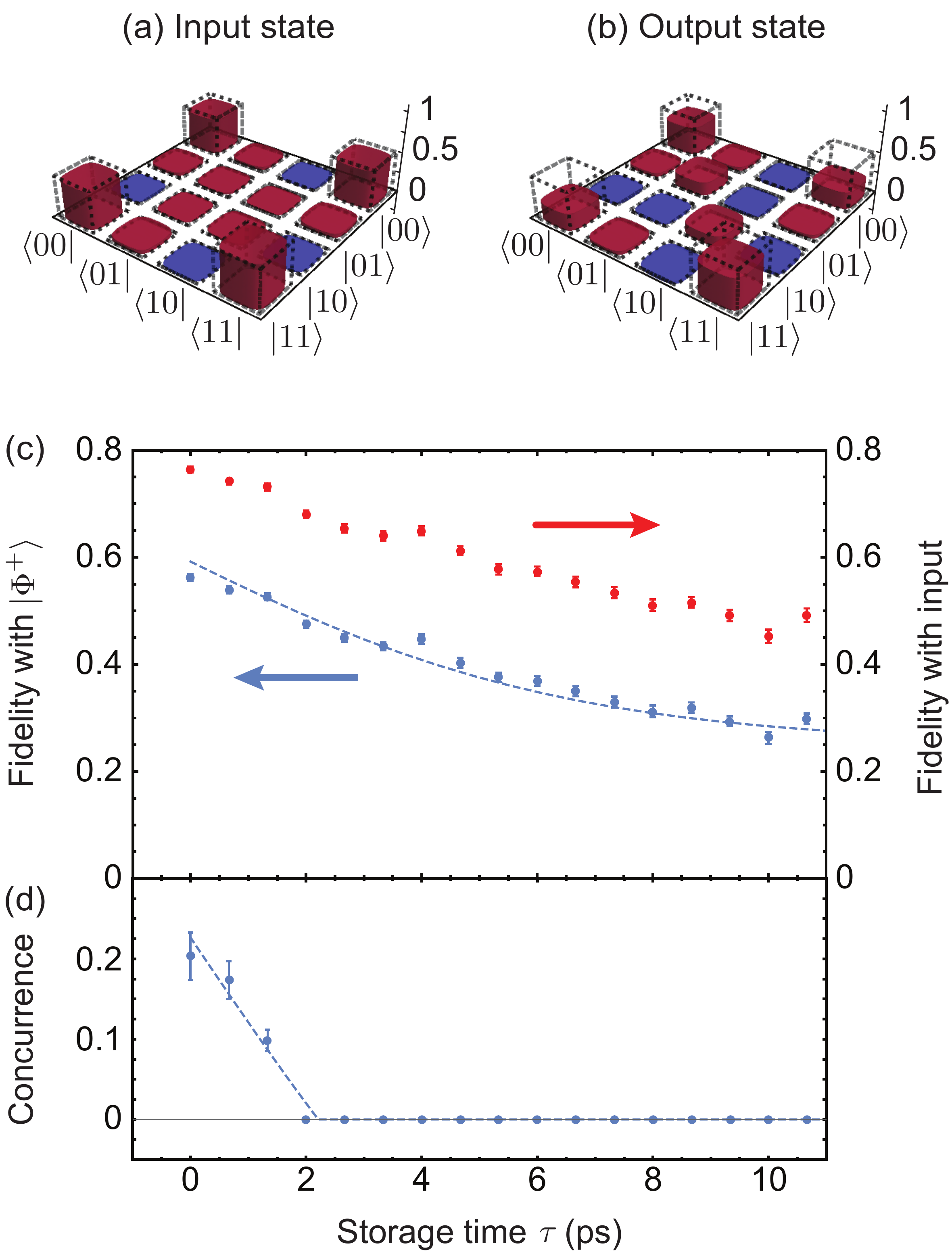}}
\caption{ Real parts of the reconstructed input (a) and output (b) two-qubit density matrices. The input state has $0.939\pm 0.001$ fidelity with $\ket{\Phi^+}$, whereas the state output from the quantum memory has $0.562 \pm 0.007$ fidelity with $\ket{\Phi^+}$ and $0.764 \pm 0.005$ with the input state. (c) Fidelity of the output state with $\ket{\Phi^+}$ (blue dots, left axis), and with the input state (red dots, right axis), as a function of storage time. (d) Concurrence of the output state (dots) and for a Werner state (dashed) as a function of storage time. Entanglement persists for up to 1.3\,ps of storage.     }
\label{fig:Tangle}
\end{figure}

%%% ENTANGLEMENT %%%
We now turn to the storage of one photon of an entangled pair. To prepare a two-photon polarization-entangled state we pump type-I \SPDC in two orthogonally oriented 1\,mm \BBO crystals using a diagonally-polarized pump pulse. With appropriate temporal compensation and phase correction, this ideally generates the state $\ket{\Phi^+} = ( \ket{0}_{s}\ket{0}_{h} +  \ket{1}_{s}\ket{1}_{h})/ \sqrt{2}$ where subscripts $s$ and $h$ refer to signal and herald photons, respectively. In order to store the entangled signal photon we remove the state preparation apparatus from the signal path shown in Figure~\ref{fig:Concept}(a); otherwise the procedure is identical to above. Using a two-qubit maximum likelihood state tomography~\cite{James2001} we reconstruct the density matrix of the retrieved signal-herald state. We measure the initial state, before storage, to have $0.939\pm 0.001$ fidelity with the $\ket{\Phi^+}$ state; the real part of the input state density matrix is shown in Figure~\ref{fig:Tangle}(a). We measure an input coincidence rate of $\sim$360\,Hz in both $\ket{0}_{s}\ket{0}_{h}$ and $\ket{1}_{s}\ket{1}_{h}$ bases, and, upon retrieval from the memory, we find a coincidence rate of $\sim$2.3\,Hz in each basis.  The reconstructed density matrix for the retrieved state is shown in Figure~\ref{fig:Tangle}(b). We find at the peak that the retrieved state has $0.764 \pm 0.005$ fidelity with the input, and $0.562 \pm 0.007$ fidelity with $\ket{\Phi^+}$. These fidelities as a function of storage time are shown in Figure~\ref{fig:Tangle}(c).  

When the process in Eq.~\ref{eq:channel} acts on one qubit from the state $\ket{\Phi^+}$ the result is a Werner state $\rho_\text{W}(\tau)=p(\tau) \ket{\Phi^+}\bra{\Phi^+} + [1-p(\tau)] \openone_4 / 4$. The fidelity of $\rho_\text{W}(\tau)$ with the state $\ket{\Phi^+}$ is $[1 + 3p(\tau)]/4$. Note that our measured state fidelity implies that $p(0) = 0.416$ which is above the classical limit of $p=1/3$.  The fidelity of $\rho_\text{W}(\tau)$ and $\ket{\Phi^+}$ is plotted in Figure~\ref{fig:Tangle}(c) (dashed line) where $S_0 = 1.39 \pm 0.07$\,Hz, $N_c = 1.41 \pm 0.04$\,Hz and $N_0 = 0.46 \pm 0.09$\,Hz. 
%Over all investigated storage times output states show high fidelity with $\rho_\text{W}(\tau)$, averaging to $0.990 \pm 0.001$.

%%% FIG %%%

We use concurrence~\cite{Wootters1998} to quantify the amount of entanglement in the retrieved state.  The concurrence of a state $\rho$ is $\mathcal{C}(\rho) = \max \{0, \lambda_1 -\lambda_2 -\lambda_3 -\lambda_4 \}$, where $\lambda_i$ are the eigenvalues, in decreasing order, of the matrix $\sqrt{ \sqrt{\rho } \tilde{\rho} \sqrt{\rho } }$; here $\tilde{\rho} = (\sigma_{y} \otimes \sigma_{y}) \rho^* (\sigma_{y} \otimes \sigma_{y})$. The concurrence of the input state is $0.832\pm0.003$. (Maximally entangled states have $\mathcal{C} = 1$, separable states have $\mathcal{C} = 0$.)  Figure~\ref{fig:Tangle}(d) shows the concurrence of the retrieved signal as a function of storage time. The concurrence for a Werner state, $\mathcal{C}(\rho_W (\tau)) = \max \{0, [3p(\tau)-1]/2 \}$, is plotted alongside the data (dashed line).  Though the state is diminished due to noise, entanglement between the stored and herald photons persists for up to 1.3\,ps of storage time.

As a final note, we return to the result in Fig.~\ref{fig:AvgFid}(a), which shows the retrieved signal and associated noise. Noise photons are generated by one of two processes: four-wave mixing (4WM) between read, write and Stokes pulses~\cite{England2013}; and read pulses scattering off of thermally populated phonons~\cite{England2015}. Thermal noise will be generated at a constant rate of $N_\text{th}$. The four-wave mixing noise contains both two-pulse (read and write) and single-pulse (read only) contributions so the delay-dependent four-wave mixing rate is given by $N_\text{4WM}  ( 1 + e^{-\tau / \tau_\text{m}} )$. We use the experimentally measured constant ($N_c$) and time-dependent ($N_0$) noise rates to extract the thermal and four-wave mixing noise rates as $N_\text{th} = N_c-N_0 $ and $N_\text{4WM} = N_0$. We therefore see that around 80\% of the noise in this experiment is thermal, which could be mitigated by cooling. For instance at $-40^\circ$C, achievable with Peltier cooling, thermal noise should be reduced by a factor of 6.  From our model we expect that the peak average fidelity (Fig.\ref{fig:AvgFid}(b)) would increase to 0.91 and would remain above the classical bound for 7.6\,ps of storage, over 28 times longer than the input photon. The entangled state fidelity with $\ket{\Phi^+}$  would increase to 0.787 and the concurrence would remain above zero for 6\,ps of storage. As seen in Ref.~\cite{England2013}, memory efficiency scales quadratically with control pulse energies up to 10\,nJ. However, the noise produced will also increase: four-wave mixing scales quadractically, while thermal noise depends only on the read pulse and so scales linearly. We estimate that read and write pulse energies of 10\,nJ would increase the memory efficiency  to $\sim$4\% and average fidelity to 0.86.
%this is calculated using the equation for $p(\tau)$, rewriting the noise in terms of N_th and N_4wm, adding a quadratic coefficient (x^2) to S_0, N_4wm and a linear coefficient (x) to N_th.  If x = 10 / 4.4  we get p(\tau=0) = 0.7236 which gives F_avg = 0.862

%%% Conclusion %%%
We have shown that the diamond memory supports on-demand, non-classical storage of entangled photonic qubits with THz bandwidths.   While the storage time is too short for many long-distance communications, we expect that the diamond memory will find application where long-lived storage is not required, but instead only a large time-bandwidth product, i.e., the number of operational time-bins a photon is stored for.  This includes enhancing multi-photon rates~\cite{Nunn2013} and processing spectral properties of photonic qubits~\cite{Fisher2016, Bustard2017}.  Furthermore, our present result serves as a proof-of-principle for all Raman memories, some of which have more promising storage times~\cite{Reim2010, Bustard2013}. In our demonstration we have successfully used spatial multiplexing for polarization entanglement storage. This can also be applied to the frequency degree of freedom for use in for example, rapidly reconfigurable quantum logic gates~\cite{Campbell2014},  or constructing large time-frequency cluster states~\cite{Humphreys2014}, where entanglement storage is a requirement. Extending the capacity of the diamond processing platform to include entanglement storage will make it an increasingly valuable tool for high-bandwidth quantum information applications.

%%%%%%%%%%%%%%%%%%%%%%%%%%%
\begin{acknowledgments}
This work was supported by the Natural Sciences and Engineering Research Council of Canada, Canada Research Chairs, Canada Foundation for Innovation, Industry Canada, Ontario Centres of Excellence, and Ontario Ministry of Research and Innovation Early Researcher Award. The authors are grateful to Matthew Markham and Alastair Stacey of Element Six Ltd. for manufacturing the diamond sample, to Doug Moffatt for software assistance, and to Denis Guay for technical support. They also acknowledge fruitful discussions with Rune Lausten, Paul Hockett, John Donohue, Michael Mazurek and Josh Nunn.
\end{acknowledgments}

\bibliography{qubitstorage}

\begin{thebibliography}{34}
\bibitem[{\citenamefont{{The process fidelity of two process matrices, $\chi_1$
  and $\chi_2$, is calculated by $\mathcal{F}_p(\chi_1, \chi_2)=\left (
  \text{Tr} \sqrt{ \sqrt{\chi_2} \cdot \chi_1 \cdot \sqrt{\chi_2} } \right
  )^2$}}()}]{procfid}
\bibinfo{author}{\bibnamefont{{The process fidelity of two process matrices,
  $\chi_1$ and $\chi_2$, is calculated by $\mathcal{F}_p(\chi_1, \chi_2)=\left
  ( \text{Tr} \sqrt{ \sqrt{\chi_2} \cdot \chi_1 \cdot \sqrt{\chi_2} } \right
  )^2$}}}.
\end{thebibliography}

\end{document}